\documentstyle[12pt,aaspp]{article}

\def\pz{\phantom{0}}
\def\lsim{\mathrel{\lower .85ex\hbox{\rlap{$\sim$}\raise
.95ex\hbox{$<$} }}}
\def\gsim{\mathrel{\lower .80ex\hbox{\rlap{$\sim$}\raise
.90ex\hbox{$>$} }}}

\begin{document}

\def\thefootnote{\fnsymbol{footnote}}

\title{Dwarf Cepheids in the Carina Dwarf Spheroidal Galaxy
\footnote{Based on observations obtained at CTIO}}

\def\thefootnote{\arabic{footnote}}

\author{Mario Mateo\altaffilmark{1}, Denise Hurley-Keller\altaffilmark{1}}
\affil{\tt e-mail: (mateo,denise)@astro.lsa.umich.edu}
\author{James Nemec\altaffilmark{2}}
\affil{\tt e-mail: jmn@isr.bc.ca}

\altaffiltext{1}{Department of Astronomy, University of Michigan, 821 Dennison 
Bldg., Ann Arbor, MI \ \ 48109--1090}
\altaffiltext{2}{International Statistics and Research Corp., P.O. Box 496,
Brentwood Bay, BC \ \ V8M 1R3}

\begin{abstract}

We have discovered 20 dwarf Cepheids (DC) in the Carina dSph galaxy
from the analysis of individual CCD images obtained for a deep
photometric study of the system.  These short-period pulsating
variable stars are by far the most distant ($\sim$100 kpc) and
faintest ($V \sim 23.0$) DCs known.  The Carina DCs obey a
well-defined period-luminosity relation, allowing us to readily
distinguish between overtone and fundamental pulsators in nearly every
case.  Unlike RR~Lyr stars, the pulsation mode turns out to be
uncorrelated with light-curve shape, nor do the overtone pulsators
tend towards shorter periods compared to the fundamental pulsators.
Using the period-luminosity (PL) relations from Nemec et al. (1994 AJ,
108, 222) and McNamara (1995, AJ, 109, 1751), we derive $(m-M)_0 =
20.06 \pm 0.12$, for E$(B-V)$ = 0.025 and [Fe/H] = $-$2.0, in good
agreement with recent, independent estimates of the distance/reddening
of Carina.  The error reflects the uncertainties in the DC distance
scale, and in the metallicity and reddening of Carina.  The frequency
of DCs among upper main sequence stars in Carina is approximately
3\%. The ratio of dwarf Cepheids to RR~Lyr stars in Carina is 0.13
$\pm$ 0.10, though this result is highly sensitive to the
star-formation history of Carina and the evolution of the Horizontal
Branch.  We discuss how DCs may be useful to search effectively for
substructure in the Galactic halo out to Galactocentric distances of
$\lsim 100$ kpc.

\end{abstract}

\clearpage

\section{Introduction}

The Carina dwarf spheroidal (dSph) galaxy has an important place in
the history of our understanding of stellar populations in external
galaxies.  At the time of its discovery (Cannon et al. 1977), there
were six known dSph satellites of the Milky Way (for an up-to-date
census of these systems in the Local Group see Da Costa 1998 and Mateo
1998a,b).  All were believed to be essentially old stellar systems
similar in stellar population to globular clusters.  Some evidence to
the contrary did exist: Baade (1963) and later Zinn (1980) emphasized
that the variable stars of dSph galaxies -- in particular the RR~Lyr
stars and Anomalous Cepheids -- suggested that their underlying
stellar populations were fundamentally different from those of
globular clusters.  Zinn postulated that the difference was age, such
that the dSph systems were on average younger than globulars.
Subsequent deep photometric observations of the main sequence turnoffs
(MSTO) in these galaxies confirmed this conjecture so spectacularly
that it is now difficult to identify {\it any} dSph galaxy whose
population is as uniformly old as that of a typical globular cluster
(see Da Costa 1998 for an excellent recent review)!  Carina was the
first dSph galaxy so observed in the early days of modern CCD
detectors (Mould and Aaronson 1983) and remains one of the more
spectacular examples of a dSph galaxy with a significant
intermediate-age population.  Mighell (1990, 1997), Smecker-Hane et
al.  (1994a) and Hurley-Keller et al. (1998; hereafter Paper~I) have
obtained additional photometry of the Carina main-sequence that
confirms the bulk of its stars have ages ranging from about 4-7 Gyr,
though older (Saha et al. 1986) and younger stars are also clearly
present.

One consequence of the relative youth of Carina is that its main
sequence extends far above the canonical MSTO luminosity of globular
clusters.  Many main-sequence stars are found in Carina within the
instability strip where short-period pulsating stars known as dwarf
Cepheids, SX~Phe stars, $\delta$~Sct stars, or ultra-short-period
Cepheids are found (see Mateo 1993 for a review of the properties and
confused nomenclature of these stars).  In this paper we shall refer
to these variables as `dwarf Cepheids' (DC) to emphasize that they are
main-sequence stars pulsating in basically the same manner as
classical Cepheids.  Nemec et al. (1994; hereafter NNL) provide a
comprehensive survey of the then-known Population II dwarf Cepheids,
referring to these stars as `SX~Phe' variables.  Many DCs also happen
to be `blue stragglers' in their parent clusters, located above and to
the blue of the MSTO.  These blue-straggler sequences sometimes extend
to the dwarf Cepheid instability strip, a region normally forbidden to
stars older than about 8-10 Gyr (see Figure 1).  Carina is a logical
place to search for DCs as its main sequence extends up to and through
the region where dwarf Cepheids are found in other stellar systems
as a result of its prominent intermediate-age population (Figure 1).

In this paper we describe a search for dwarf Cepheids in the Carina
dSph galaxy.  The data used here are the same as those reported in
Paper~I, except that now we have analyzed the frames individually to
perform time-series photometry, rather than combining the images to go
as deep as possible.  Two challenges face us in this search.  First,
DCs have very short periods ranging from about 1 hour to 3 hours.  Any
exposures longer than 10-15 minutes can `smear out' these rapid light
variations.  Second, because DCs have modest luminosities ($M_V \sim
+3$ to $+1$), they are expected to have $V$ magnitudes of $\sim$22-24
in Carina, rather faint for such short individual exposures.
Nonetheless, as we describe here, we have discovered 20 DCs in Carina.

\section{Observations and Reductions}

\subsection{General Procedures}

The observations used here are the same as reported in Paper~I.  To
review, the data were obtained using the CTIO 4m telescope with the
prime-focus CCD camera during two runs in 1989 and 1990.  At the time,
this camera contained a thinned, backside-illuminated TI $800 \times
800$ CCD.  The typical exposure time was 500~s in both the $V$ and $B$
filters.  A log of the individual exposures is given in Table 1.
Three fields were observed, and a chart showing their relative
locations is given in Paper~I.  Field 1 was observed only in 1989, and
fields 2 and 3 were observed in 1990.  A known flat-fielding defect
afflicts the first 100 columns of the CCD used in both runs; no data
from this section of the CCD were used in the analyses here or in
Paper~I.  Further details of the observations can be found in Paper~I.

The individual processed frames were measured using the DoPHOT
photometry program (Schechter et al. 1993) in the same manner as in
Paper~I.  Some of the frames with slightly poorer seeing than average
that were excluded from the analysis of Paper~I, but they are used
here because we could reject individual poor measurements at later
stages as required.  Details of the calibration of the resulting light
curves are described below, but these are based on the
transformations described in Paper~I.

\subsection{Identification of Variable Stars}

Once the individual CCD images were reduced with DoPHOT, we registered
the coordinates to a common system.  The typical rms of these
transformations was 0.07 pixels, or about 0.04 arcsec.  The data for a
given field and filter were then combined into databases in which the
individual photometric results were tabulated for each object found on
every frame.  Approximately 800-1000 stars per field were found in
each exposure compared to the 2000-3000 stars found in each of the
deep images of each field in Paper~I.  There are fewer total epochs
for Fields 2 and 3 compared to Field 1, but all three fields were
observed on at least two separate nights within their respective runs.

The photometric results for each image were then shifted to a common
system defined by an arbitrarily-chosen template image.  After some
experimentation, we found that a simple additive shift to the
instrumental $V$ and $B$ magnitudes was sufficient to bring all the
images of a given filter and field to a common zero-point.  We then
determined the photometric transformation of the template images using
the calibrated data from Paper~I for each field.  This transformation
included the mean color terms found for the three fields, as well as a
frame-specific offset used to convert to the standard system. For
reasons we describe later, these transformations were not applied
to the database photometry at this point.

Instead, we first identified candidate variables from the
pre-calibrated but photometrically-shifted results.  A first cut was
made by calculating $\chi^2$ under the hypothesis that the stars are
non-variable.  Although pulsating variables tend to have larger
amplitudes in $B$ than in $V$, we determined $\chi^2$ from the
$V$-band data because we obtained many more epochs in that filter.
Stars with `large' values of $\chi^2$ were then extracted from the
databases.  We chose to be rather conservative at this stage, and
selected a value of $\chi^2$ that allowed us to extract about 10\%\ of
the total number of stars in the databases for each field.  Our
motivation was to try to ensure complete coverage of all variables at
the cost of having to sort through a significant number of
non-variables in the process.  At this point we also required that any
candidate variable must have good photometric results from at least
90\%\ of the individual images.  This was particularly important for
Fields 2 and 3 which have relatively few observations; any fewer would
have made it impossible to identify variables reliably.  About 50-80
variable-star candidates were identified in each field with this
procedure.

We then took the photometry of the candidates and did a period search
using the string algorithm described by Dworetsky (1983).  The nine
best periods were then used to fold the data and the resulting phased
light curves were plotted together and inspected.  The initial period
search was rather broad, ranging from 0.05 to 2.0 days.  In this
manner we isolated 25 candidate DCs in the three Carina fields.  We
then turned to each individual star and determined a `best' period
using the same algorithm, but this time searching only in the range
0.05-0.10 days.  Because of our limited time coverage, the correct
period is not always obvious; we estimate an uncertainty of up to
10\%\ in the worst cases.  For most of the stars, the periods we
derived are the only ones consistent with the photometry and are
probably accurate to the stated precision.  When there was some
ambiguity about the period, we selected the shortest period consistent
with the data in the final period range noted above.  This procedure
probably biases our most uncertain periods to small values.  Because
every field was observed on multiple nights, we deleted some promising
candidates that did not repeat consistently even when their phased
light curves looked quite convincing.

The phased $V$-band light curves were used to determine the
ephemerides for each candidate; these are listed in Table~2 in the
form of the period (in days) and the approximate time of minimum
light.  Because of the long interval between the time the data were
taken and the present, these quantities are likely to be of little use
to phase newer data.  We have included these ephemerides in case some
other data more contemporaneous to those reported here were obtained
and are eventually analyzed.  The $B$-band light curves were phased
using these ephemerides.  In a few remaining cases these were found to
be inconsistent with the $V$-band data, either in the phasing or
because the $B$-band data did not confirm variability.  Such stars
were rejected as candidate variables.

We then calibrated the light curves of the remaining candidates as
follows.  Because the transformations to the standard system involve
modest color terms, we could not simply transform the $V$ and $B$
photometry independently.  Also, because DCs vary in color during
their pulsation cycle, we could not adopt a mean color.  To deal with
this we took one of the phased light curves (say, $V$) for a given
star, and for each point calculated the corresponding instrumental
magnitude at the same phase from the phased light curve in the other
filter (in this case $B$).  This was done by linearly interpolating
the second light curve about the the desired phase from the first
light curve.  This procedure was then repeated for the other light
curve (in this case $B$).

\section{Dwarf Cepheids in Carina}

At the end of this process we identified 20 dwarf Cepheids in our
three fields; 12 in Field~1, and four each in Fields 2 and 3.  A
summary of the properties of these 20 variables is provided in Table
2, and their light curves are shown in Figure 2.  We have named the
stars V1-20.  The individual photometry results for every epoch and
every faint variable is listed in Table 3.  Note that the mean $B$ and
$V$ magnitudes and mean $B-V$ colors for the variables are taken from
Paper~I.  Particularly for the $B$ band, we felt it was more reliable
to derive the averaged photometry from the co-added images than from
the light curves.  Only two stars, V11 and V12, were not found in the
photometry list from our deep images from Paper~I.  In these cases the
mean photometry values in Table 2 are obtained from intensity averaged
means of the $B$ and $V$ light curves.  Note that the colors
listed in Table 2 are strictly defined as $\langle B\rangle - \langle
V\rangle$, and this is true of V11 and V12 also.  The locations of the
Carina DCs is shown in Figure 3.  The coordinates of the centers of
the three fields are listed in the caption to Figure 3.
Further details can be found in Paper~I. 

Given the quality of the light curves of the faintest variables (V7
and V8), we estimate that out procedure was capable of finding
variables in this data set to $V \sim 23.3$, or about $M_V \sim 3.1$
in Carina.  This limit is about 2 magnitudes brighter than the
detection limit of our deep combined data, and may serve as a figure
of merit for any future surveys that obtain sufficient exposures
($\gsim 10$) to identify such faint, short-period variables.  We have
not attempted to determine the completeness of our data as a function
of magnitude, location, or photometric amplitude; our results
represent a lower limit to the number of DCs in Carina.

The locations of the variables in the color-magnitude diagram of
Carina is shown in Figure 4 using the mean photometry from Table 2.
Two features are apparent from this diagram.  First, the DCs do indeed
cluster near the upper MS; we shall discuss this further below.
Second, the approximate completeness limit for the detection of DCs in
these data is well below the mean brightness of the variables we did
identify.  {\it Many} stars above this limit and in the same general
location as the DCs are definitely not variable with amplitudes larger
than 0.02-0.04 mag in the $V$ band.  We shall also return to the
implications of this point below.

\section{Discussion}

\subsection{The Period-Luminosity Relation(s) of the Carina Dwarf Cepheids}

We have plotted in Figure 5 the period-apparent magnitude data for the
dwarf Cepheids in Carina.  Plotted as solid lines are the fundamental
and first-overtone dwarf Cepheids from NNL (they refer to them as
SX~Phe stars in this paper) for an apparent $V$-band distance modulus
of 20.17 and a mean metallicity of $-2.0 \pm 0.2$ using their $V$-band
period-luminosity-metallicity relations for these stars.  This modulus
is taken from Smecker-Hane et al. (1994b) who estimate a mean
reddening of E$(B-V)$ = 0.025.  We have also adopted the metallicity
proposed by these authors that they claim applies to the younger stars
in Carina.  This value also agrees with the latest spectroscopic
estimate of Da~Costa (1994).  We need to know the abundance for this
analysis because NNL explicitly include a metallicity term in their
period-luminosity relations for DCs.

Figure 5 shows that the Carina DCs split into two groups that obey
distinct period-luminosity (PL) relations.  The brighter group
consists of first-overtone pulsators, while the fainter group contains
fundamental pulsators.  The principal uncertainties in the data
plotted in Figure 5 are -- uncharacteristically -- in the horizontal
axis, i.e. in the the periods.  However, it is important to note that
no stars can reasonably be switched from the overtone to fundamental
group (or {\it vice versa}) due to period uncertainties, apart from
V2.  From Figure 5 it is also clear that the period range of the
fundamental pulsators overlaps that of the overtone variables.  Unlike
RR~Lyr stars, short-period DCs are {\it not} predominately overtone
pulsators. Inspection of Figure 2 also reveals that the light curve
shape is not uniquely correlated with the pulsation mode (which is
noted for each variable in the figure).  Thus, one cannot simply
assume that sinusoidal light curves correspond to overtone pulsators
(see V5), nor that a `sawtooth' light curve identifies a
fundamental-mode pulsator (see V12 and V18).  Because mass
determinations for dwarf Cepheids require knowledge of the pulsation
mode (e.g. see Da Costa et al. 1986), some caution is needed to assign
the mode.

Figure 5 also reveals very good agreement between the Carina DCs and
the NNL period-luminosity relation when we adopt the Smecker-Hane et
al. (1994b) distance, metallicity and reddening noted above.  For
these parameters the data agree well with the slope of the PL relation
from NNL, particularly for the fundamental-mode pulsators which
exhibit a wider period range than the overtone pulsators.  The data
are also consistent with the relative separation of the overtone and
fundamental PL relations.   All of these points suggest strongly that
the Carina DCs are `normal' for their class, despite their possibly
different evolutionary origin (Mateo 1993).

McNamara (1995) derived a PL relation for DCs and used preliminary
results for the Carina DCs (Nemec and Mateo 1990) to derive a distance
modulus of $20.01 \pm 0.05$ and [Fe/H] = $-1.5$, for E(B$-$V) = 0.025.
For [Fe/H] = $-2.0$, the McNamara (1995) PL relation gives a true
modulus of $20.06 \pm 0.05$.  We can repeat this analysis using the
most recent (and reliable) data on the Carina DCs, and incorporating
some other constraints on the metallicity and reddening of the
galaxy. For example, if we fix the slope of the PL relations and set
the metallicity dependence to the values suggested by NNL, we find a
best estimate for the true distance modulus to be 20.07 $\pm$ 0.03 for
the metallicity and reddening adopted above.  This result agrees well
with the distance found by Smecker-Hane et al. (1994b). More recently,
Mighell (1997) has proposed a true modulus of 19.87 $\pm$ 0.11 for
Carina, with a visual extinction of 0.18 (E$(B-V)$ = 0.051), and
[Fe/H] = $-$1.9.  A comparison of the resulting PL relations for these
parameters is shown in Figure 6.  There is a small systematic offset
from the data for the Carina variables (0.09 $\pm$ 0.03 mag); our best
estimate of the true modulus is 19.96 $\pm$ 0.03 for this reddening
and metallicity.

None of these results include realistic estimates of the uncertainties
in the DC distance scale adopted by NNL which is tied directly to the
distance of the metal-poor globular cluster M~15.  NNL adopted a true
modulus of 15.03; for comparison, Durrell and Harris (1993) derive a
true modulus of $15.09 \pm 0.15$ for the cluster. This change results
in a Carina modulus of 20.13 and 20.02 for the Smecker-Hane et
al. (1994b) and Mighell (1997) reddening/metallicity parameters,
respectively.  The uncertainties of the distance scale adopted by
McNamara (1995) are difficult to quantify because the calibrating
stars are taken from a number of different sources and clusters; it is
likely that this scale is also uncertain at the 10-15\%\ level.  We
conclude that the best estimate of the distance to Carina is
approximately $20.05 \pm 0.12$ based on the DCs alone and assuming
[Fe/H] = $-2.0$ and E(B$-$V) = 0.025.

\subsection{The Frequency of Dwarf Cepheids in Carina}

Figure 7 shows a region enclosing the upper portion of the main
sequence in Carina and all 20 newly-discovered dwarf Cepheids.  We
know from Paper~I that the completeness in this region is quite high
and uniform, and that the total field contamination is quite small.
Thus, we have simply counted the total number of stars in this region
(844) to estimate the frequency of DCs in Carina.  We find that
approximately one of 42 stars in this region of the color-magnitude
diagram is a dwarf Cepheid, or about 2.5\%\ of all upper main-sequence
stars in Carina.  This is a lower limit because we have probably
missed some small-amplitude variables.  

One consequence of this result has to do with the suitability of DCs
as tracers of the sort of stellar population present in Carina.  There
are numerous hints suggesting that the halo has considerable
substructure, possibly from the accretion of small dwarfs (Rodgers et
al. 1981; Majewski 1992; C\^ot\'e et al. 1993; Arnold and Gilmore
1992; Preston et al. 1994).  The discovery of the Sgr dwarf galaxy
(Ibata et al. 1994) is a spectacular example that such accretion does
occur even now to some extent.  Systematically identifying and tracing
the substructure has proven to be quite difficult.  Can DCs help?

In the case of Carina, the total area we surveyed for DCs is 0.013
deg$^2$, implying a mean surface density of DCs of 1540 $\pm$ 340
deg$^{-2}$ (Poisson error).  The mean $V$-band surface brightness of
Carina in these fields is about 25.5 mag arcsec$^{-2}$.  If instead we
had viewed the same stellar population found in Carina in a structure
spread out to an average $V$-band surface brightness of 30.5 mag
arcsec$^{-1}$, we would have found approximately 15 DCs per square
degree.  Such a faint surface brightness is unobservable in any
conventional sense, but the detection of the associated DCs {\it is}
possible.  A deep, wide-field search for DCs might be useful to
uncover extended structure beyond the tidal radius of Carina, or
perhaps other extended structures with similar populations elsewhere
in the halo.  Closer halo structures would be more spread out,
consequently requiring less depth but greater areal coverage to
identify the dwarf Cepheids.

\subsection{The Dwarf Cepheid Instability Strip}

In Figure 7 we plot the boundaries of the DC instability strip (IS)
from Breger (1979) and from Mateo (1993) superimposed on the CMD of
Carina.  The locations of the Carina DCs are consistent with the
Breger boundaries.  Most of the DCs lie within the more confined
boundaries from Mateo which was based on the locations of these
variables in other intermediate-age and old stellar systems (in the
latter case, globular clusters with blue stragglers).  The Breger IS
boundaries are based on the locations of DCs associated with younger
populations, and in particular the so-called $\delta$-Sct stars which
generally have very low amplitudes.  These stars are generally drawn
from field samples; random errors in their reddenings and distances
contribute to uncertainties in the location of the instability strip
edges.  Because of the limited $B$-band phase coverage for the Carina
variables, the mean colors are sometimes in error by up to 0.1-0.2
magnitudes.  This introduces some of the scatter visible in Figure 7.
The close conformance of the Carina DCs to the IS boundaries is
further evidence that these variables are normal for their class.

An obvious feature of Figure 7 -- basically a corollary of the point
raised in section 4.2 above -- is that not all stars within the
dwarf-Cepheid instability strip are actually dwarf Cepheids.  We have
looked at the light curves of a number of these nonvariables, and they
are certainly constant to the precision of the data, or about
0.03-0.05 mag at the mean magnitude range of the detected DCs.  Among
RR~Lyr stars and classical and Type~II Cepheids, it is generally
agreed that {\it any} star in the IS will begin to pulsate and be seen
as the appropriate sort of variable.  This is clearly not true of
large-amplitude DCs where we find that only 2-3\%\ of the stars in the
IS are variable with V-band amplitudes larger than 0.03-0.05 mag.

\subsection{Other Variable Stars in the Fields}

Saha et al. (1986; hereafter SSM) first searched for variable stars in
Carina using photographic plates with the CTIO 4m telescope.  Their
search targeted RR~Lyr stars and Anomalous Cepheids in Carina and was
not deep enough to detect the DCs described here.  Our data do include
some measurements of the RR~Lyr stars and candidates identified by
SSM. A summary of the results of our observations of these stars is
provided in Table 4.  We confirm the two RR~Lyr stars for which SSM
derived periods and light curves: SSM~74 and SSM~116.  A third
variable -- SSM~65 -- was not included in our dataset due to its
proximity to the central bright star in Field~3.  Of the 10 suspected
variables from SSM that have adequate data in our observations we find
that nine are certainly non-variable to $\pm 0.02$ mag.  For the tenth
star -- SSM~87 -- our data cannot confirm variability because the star
is too close to the edge of Field~2.  We also find one candidate
RR~Lyr variable that was not identified by SSM which we call SSM~176
to continue SSM's numbering sequence.  The locations in the
color-magnitude diagram of the two well-observed RR~Lyr stars from SSM
and our two additional candidate variables are shown in Figure 4.
Their locations in each Carina field are noted in Figure 3 along with
the locations of the other apparently non-variables from SSM.  Our
data are inadequate to determine periods much longer than 0.25 days
without severe aliasing, so we cannot comment on the periods for
SSM~74 and 116, and we are unable to determine a reliable period for
SSM~176.  The location of SSM~176 in the CMD (Figure 4) is consistent
with its identification as a possible RR~Lyr star.

We have identified an additional variable, SSM~177, that appears to
exhibit portions of two eclipses.  We cannot determine its period with
any precision and its image appears to be composite in many of the
individual CCD frames.  We regard this star as only a weak candidate
eclipsing contact binary that clearly requires confirmation.  The
photometric data for both SSM~176 and 177 are listed in Table 4 and
their locations are marked in Figure 3.

If the numbers of RR~Lyr stars and DCs in our fields are typical of
Carina as a whole, we find the ratio of these variables to be
$N_{DC}/N_{RR} \sim 2.5/20$, or 0.13 $\pm$ 0.08 (the fraction in the
numerator represents SSM~176 which is the uncertain RR~Lyr star found
in our study).  Of course, the frequency of RR~Lyr stars is highly
variable and the numbers of DCs requires a significant
intermediate-age population or blue stragglers.  Nonetheless, this
result does suggest that DCs can greatly outnumber RR~Lyr stars in
populations of satellites in the outer Galactic halo.

\section{Conclusions}

We have discovered 20 dwarf Cepheids in the Carina dSph galaxy.  These
are by far the most distant ($\sim$100 kpc) and faintest ($V \sim
23.0$) DCs known.  The Carina DCs obey a well-defined
period-luminosity relation, allowing us to distinguish between
overtone and fundamental pulsators in nearly every case.  Unlike
RR~Lyr stars, the pulsation mode turns out to be uncorrelated with
light-curve shape, nor do the overtone pulsators tend towards shorter
periods compared to the fundamental pulsators.  Using the PL relations
from NNL, we find good agreement with the distance/metallicity
parameters advocated by Smecker-Hane et al. (1994b): $(m-M)_0 =
20.09$, E$(B-V)$ = 0.025, and [Fe/H] = $-2.0$ and for the results of
Mighell (1997) using slightly different parameters.  Our best estimate
of the Carina true distance modulus is $20.06 \pm 0.12$, where the
uncertainty includes the errors of the DC distance scale (NNL;
McNamara 1995).  This corresponds to a distance of $103 \pm 6$ kpc.
The frequency of DCs among upper main sequence stars in Carina is
approximately 3\%\ in the region enclosed in Figure 4.  We find that
the ratio of dwarf Cepheids to RR~Lyr stars in Carina is 0.13 $\pm$
0.10, though of course the numbers of such stars are highly sensitive
to the details of the star-formation history of Carina and the
evolution of the Horizontal Branch, respectively.

The fact that we could discover these faint, short-period variables
using a relatively poor CCD in mediocre seeing conditions (by modern
standards) on a 4m telescope suggests that these stars could
conceivably be used as tracers of extended structure throughout the
Galactic Halo.  About one out of 40 upper main sequence stars in
Carina is a DC.  If Carina has a tidal tail or an extended halo (see
Lynden-Bell and Lynden-Bell 1995 for some motivation why this may
occur; Kuhn et al. 1996), DCs may prove to be an excellent way to
efficiently trace such extended structures.  As noted in section 4.2,
if Carina were spread out to 1\%\ of its current surface brightness we
would expect to see about 15 of these stars per square degree.  More
generally, any survey of the Galactic halo should identify one DCs per
40 so-called blue-metal-poor (BMP) star.  These stars were first
described by Preston et al. (1994; though their lineage extends back
at least to Rodgers et al. 1981) and may represent an intermediate-age
population that comprises up to 10\%\ of the Galactic halo.  Based on
one detected DC, Preston and Landolt (1998) find a frequency of about
half this in their sample of spectroscopically studied BMP stars; this 
result is not statistically different from the frequency we find
in Carina.  In an on-going survey of the halo field, Harding,
Morrison, Olszewski and MM find approximately 15 BMP stars per square
degree to a limiting magnitude of $V \sim 20.0$.  Thus, one halo DC
should occur in every three-square degree region covered by this
survey, more frequently for deeper surveys.  Eventually, the numbers
of BMP stars and DCs will fall off towards fainter magnitudes as the
halo density drops and the frequency of all luminous halo stars
becomes too low.  Precisely at what magnitude this occurs depends on
the structural parameters of the halo.  For a `reasonable' halo one
expects the number of BMP and therefore DCs to become negligible by $V
\sim 23$ in a halo devoid of substructure.

Other dSph stars may also contain DCs; the prerequisite seems to be
the existence of a significant intermediate-age population.  Like
Carina, both Leo~I (Gallart et al. 1998) and Fornax (e.g. Beauchamp et
al. 1995; Demers et al. 1998) contain large numbers of stars that
appear to be considerably younger than globular clusters.  All of the
remaining dSph systems apart from UMi contain at least a modest
intermediate-age population (Da Costa 1998).  It is therefore very
likely that many of these systems contain DCs.  Leo~I is probably too
distant to survey for these variables from the ground, but such
observations should be feasible for Fornax.  Located about 35 kpc
further than Carina, the DCs in Fornax are likely to have $V$-band
magnitude of about 23-24.  Though challenging, such observations are
certainly possible using current CCDs on 4m-class telescopes.  Much of
what we have written about the Carina DCs is necessarily a
generalization based on the variables of only one galaxy, so it would
be of great interest to broaden our understanding of these stars by
searching in other systems.

\clearpage

%\begin{small}
\begin{planotable}{ccccccc}
\tablewidth{35pc}
\tablecaption{Log of Observations}
\tablehead{ \colhead {Date} & \colhead{HJD} & 
 \colhead{Filter} & \colhead{Field}
 & \colhead{Exp. Time} & \colhead{Airmass} & \colhead{Seeing} \cr
 & -- 2447000.0 &  &  & (seconds) &  & (arcsec)
}

\startdata
Mar 11 1989 & 596.54913 & V & 1  &  420 &  1.09 &   1.6    \cr
Mar 11 1989 & 596.55781 & B & 1  &  600 &  1.10 &   1.9    \cr
Mar 12 1989 & 597.52663 & V & 1  &  420 &  1.08 &   1.3    \cr
Mar 12 1989 & 597.53255 & V & 1  &  420 &  1.08 &   1.3    \cr
Mar 12 1989 & 597.53847 & V & 1  &  420 &  1.08 &   1.3    \cr
Mar 12 1989 & 597.54738 & V & 1  &  420 &  1.09 &   1.5    \cr
Mar 12 1989 & 597.55332 & V & 1  &  420 &  1.10 &   1.4    \cr
Mar 12 1989 & 597.55928 & V & 1  &  420 &  1.11 &   1.4    \cr
Mar 12 1989 & 597.56528 & V & 1  &  420 &  1.12 &   1.5    \cr
Mar 12 1989 & 597.57125 & V & 1  &  420 &  1.13 &   1.5    \cr
Mar 12 1989 & 597.57784 & V & 1  &  500 &  1.14 &   1.4    \cr
Mar 12 1989 & 597.58480 & B & 1  &  500 &  1.15 &   1.5    \cr
Mar 12 1989 & 597.59195 & V & 1  &  500 &  1.17 &   1.5    \cr
Mar 12 1989 & 597.59925 & B & 1  &  500 &  1.19 &   1.5    \cr
Mar 12 1989 & 597.60628 & V & 1  &  500 &  1.21 &   1.4    \cr
Mar 12 1989 & 597.61348 & B & 1  &  500 &  1.23 &   1.4    \cr
Mar 12 1989 & 597.62072 & V & 1  &  500 &  1.26 &   1.6    \cr
Mar 12 1989 & 597.62784 & B & 1  &  500 &  1.28 &   1.5    \cr
Mar 12 1989 & 597.63496 & V & 1  &  500 &  1.31 &   1.5    \cr
Mar 12 1989 & 597.64204 & B & 1  &  500 &  1.34 &   1.5    \cr
Mar 12 1989 & 597.64928 & V & 1  &  500 &  1.38 &   1.5    \cr
Mar 12 1989 & 597.65659 & B & 1  &  500 &  1.42 &   1.7    \cr
Mar 12 1989 & 597.66439 & V & 1  &  500 &  1.46 &   1.8    \cr
Mar 13 1989 & 598.50585 & B & 1  &  500 &  1.07 &   1.5    \cr
Mar 13 1989 & 598.51289 & V & 1  &  500 &  1.07 &   1.3    \cr
Mar 13 1989 & 598.51988 & B & 1  &  500 &  1.07 &   1.4    \cr
Mar 13 1989 & 598.52684 & V & 1  &  500 &  1.08 &   1.3    \cr
Mar 13 1989 & 598.53380 & B & 1  &  500 &  1.08 &   1.3    \cr
Mar 13 1989 & 598.54074 & V & 1  &  500 &  1.09 &   1.3    \cr
\tablebreak
\cr
Mar 13 1989 & 598.54775 & B & 1  &  500 &  1.10 &   1.5    \cr
Mar 13 1989 & 598.55469 & V & 1  &  500 &  1.10 &   1.4    \cr
Mar 13 1989 & 598.56190 & B & 1  &  500 &  1.12 &   1.4    \cr
Mar 13 1989 & 598.56885 & V & 1  &  500 &  1.13 &   1.5    \cr
Mar 13 1989 & 598.57597 & B & 1  &  500 &  1.14 &   1.5    \cr
Mar 13 1989 & 598.58990 & B & 1  &  500 &  1.17 &   1.4    \cr
Mar 13 1989 & 598.59690 & V & 1  &  500 &  1.19 &   1.4    \cr
Mar 13 1989 & 598.60393 & V & 1  &  500 &  1.21 &   1.6    \cr
Mar 13 1989 & 598.61083 & V & 1  &  500 &  1.23 &   1.5    \cr
Mar 13 1989 & 598.61762 & V & 1  &  500 &  1.25 &   1.5    \cr
Mar 13 1989 & 598.62483 & V & 1  &  500 &  1.28 &   1.5    \cr
Mar 13 1989 & 598.63167 & V & 1  &  500 &  1.31 &   1.4    \cr
Mar 13 1989 & 598.63844 & V & 1  &  500 &  1.34 &   1.6    \cr
Mar 13 1989 & 598.64522 & V & 1  &  500 &  1.37 &   1.5    \cr
Mar 13 1989 & 598.65202 & V & 1  &  500 &  1.41 &   1.6    \cr
Mar 13 1989 & 598.65881 & V & 1  &  500 &  1.45 &   1.5    \cr
Apr  1 1990 & 982.53487 & B & 2  &  500 &  1.17 &   1.2    \cr
Apr  1 1990 & 982.54197 & V & 2  &  500 &  1.19 &   1.2    \cr
Apr  1 1990 & 982.54906 & B & 2  &  500 &  1.21 &   1.2    \cr
Apr  1 1990 & 982.55601 & V & 2  &  500 &  1.23 &   1.2    \cr
Apr  1 1990 & 982.56295 & B & 2  &  500 &  1.26 &   1.2    \cr
Apr  1 1990 & 982.56981 & V & 2  &  500 &  1.28 &   1.2    \cr
Apr  1 1990 & 982.57685 & B & 2  &  500 &  1.31 &   1.2    \cr
Apr  1 1990 & 982.58374 & V & 2  &  500 &  1.34 &   1.2    \cr
Apr  1 1990 & 982.59077 & B & 2  &  500 &  1.38 &   1.2    \cr
Apr  1 1990 & 982.59763 & V & 2  &  500 &  1.41 &   1.3    \cr
Apr  1 1990 & 982.60461 & B & 2  &  500 &  1.45 &   1.3    \cr
Apr  1 1990 & 982.61153 & V & 2  &  500 &  1.50 &   1.2    \cr
Apr  1 1990 & 982.62557 & V & 3  &  500 &  1.60 &   1.3    \cr
\tablebreak
\cr
Apr  1 1990 & 982.63237 & V & 3  &  500 &  1.66 &   1.3    \cr
Apr  1 1990 & 982.63919 & V & 3  &  500 &  1.72 &   1.5    \cr
Apr  1 1990 & 982.64587 & V & 3  &  500 &  1.78 &   1.4    \cr
Apr  1 1990 & 982.65249 & V & 3  &  500 &  1.86 &   1.4    \cr
Apr  1 1990 & 982.65912 & V & 3  &  500 &  1.93 &   1.4    \cr
Apr  1 1990 & 982.66654 & V & 3  &  600 &  2.03 &   1.4    \cr
Apr  1 1990 & 982.67445 & V & 3  &  600 &  2.15 &   1.3    \cr
Apr  1 1990 & 982.68247 & V & 3  &  600 &  2.28 &   1.4    \cr
Apr  2 1990 & 983.50585 & V & 3  &  500 &  1.12 &   1.2    \cr
Apr  2 1990 & 983.52660 & B & 3  &  500 &  1.16 &   1.1    \cr
Apr  2 1990 & 983.53704 & V & 3  &  500 &  1.19 &   1.1    \cr
Apr  2 1990 & 983.54415 & V & 3  &  500 &  1.21 &   1.1    \cr
Apr  2 1990 & 983.55078 & V & 3  &  500 &  1.23 &   1.1    \cr
Apr  2 1990 & 983.55740 & V & 3  &  500 &  1.25 &   1.2    \cr
Apr  2 1990 & 983.56403 & V & 3  &  500 &  1.27 &   1.2    \cr
Apr  2 1990 & 983.57305 & B & 3  &  500 &  1.31 &   1.2    \cr
Apr  2 1990 & 983.57987 & B & 3  &  500 &  1.34 &   1.3    \cr
Apr  2 1990 & 983.58656 & B & 3  &  500 &  1.37 &   1.3    \cr
Apr  2 1990 & 983.59320 & B & 3  &  500 &  1.40 &   1.3    \cr
Apr  2 1990 & 983.59982 & B & 3  &  500 &  1.44 &   1.4    \cr
Apr  2 1990 & 983.61250 & V & 2  &  500 &  1.52 &   1.4    \cr
Apr  2 1990 & 983.61920 & V & 2  &  500 &  1.57 &   1.3    \cr
Apr  2 1990 & 983.62591 & V & 2  &  500 &  1.62 &   1.5    \cr
Apr  2 1990 & 983.63258 & V & 2  &  500 &  1.68 &   1.6    \cr
Apr  2 1990 & 983.64425 & V & 2  &  500 &  1.80 &   1.2    \cr
vApr  2 1990 & 983.65091 & V & 2  &  500 &  1.87 &   1.2    \cr
Apr  2 1990 & 983.65755 & V & 2  &  500 &  1.95 &   1.3    \cr
Apr  2 1990 & 983.66424 & V & 2  &  500 &  2.04 &   1.4    \cr
\end{planotable}

\clearpage

{\small
\begin{planotable}{ccccccccccc}
\tablewidth{41pc}
\tablecaption{Properties of Newly-Discovered Variable Stars in Carina}
\tablehead{
\colhead{V} & \colhead{Fld} & $\alpha_{1950.0}$ & $\delta_{1950.0}$ & 
 \colhead{$\langle V\rangle$} & 
 \colhead{$\langle B\rangle$} & \colhead{$\langle B\rangle - \langle V\rangle$}
 & \colhead{P} & \colhead{E$_0$} & \colhead{$\Delta V$} & \colhead{Md} \cr 
 & & & & & & & (days) & -- 2447000.0 & & }
\startdata
  \pz1 &  1 &   6 39 50.0 & $-$50 58 55 & 22.70 &  22.97 & 0.27 &  0.05960 &  597.4661 &  0.30 &   O \cr
  \pz2 &  1 &   6 39 57.1 & $-$50 58 21 & 23.10 &  23.43 & 0.33 &  0.04792 &  597.4666 &  0.45 &   ? \cr
  \pz3 &  1 &   6 39 51.6 & $-$50 58 16 & 22.92 &  23.21 & 0.29 &  0.07697 &  597.4565 &  0.30 &   F \cr
  \pz4 &  1 &   6 39 43.1 & $-$50 57 46 & 23.02 &  23.48 & 0.21 &  0.05991 &  597.4537 &  0.45 &   F \cr
  \pz5 &  1 &   6 39 49.9 & $-$50 57 28 & 23.06 &  23.34 & 0.28 &  0.05910 &  597.4596 &  0.50 &   F \cr
  \pz6 &  1 &   6 39 45.1 & $-$50 57 15 & 22.79 &  23.10 & 0.31 &  0.07004 &  597.4868 &  0.35 &   F \cr
  \pz7 &  1 &   6 39 50.2 & $-$50 56 44 & 23.20 &  23.53 & 0.33 &  0.04872 &  597.4875 &  0.45 &   F \cr
  \pz8 &  1 &   6 39 43.7 & $-$50 56 44 & 23.23 &  23.44 & 0.21 &  0.05275 &  597.4465 &  0.50 &   F \cr
  \pz9 &  1 &   6 40 01.6 & $-$50 56 35 & 22.79 &  23.10 & 0.31 &  0.07542 &  597.4659 &  0.45 &   F \cr
 10 &  1 &  6 39 54.2 & $-$50 56 22 & 23.03 &  23.37 & 0.34 &  0.05899 &  597.4725 &  0.45 &   F \cr
 11 &  1 &  6 39 49.4 & $-$50 55 59 & 22.75 &  23.00 & 0.25 &  0.07674 &  597.4478 &  0.50 &   F \cr
 12 &  1 &  6 39 50.2 & $-$50 55 45 & 22.83 &  22.96 & 0.12 &  0.05851 &  597.4562 &  0.35 &   O \cr
 13 &  2 &  6 40 16.8 & $-$50 54 40 & 22.93 &  23.12 & 0.19 &  0.06716 &  982.4454 &  0.60 &   F \cr
 14 &  2 &  6 40 05.4 & $-$50 53 42 & 22.45 &  22.66 & 0.21 &  0.0626\pz &   982.4820 &  0.85 &   O \cr
 15 &  2 &  6 40 03.2 & $-$50 52 45 & 22.97 &  23.29 & 0.32 &  0.0578\pz &   982.4359 &  0.80 &   F \cr
 16 &  2 &  6 39 56.0 & $-$50 51 29 & 22.81 &  23.05 & 0.24 &  0.065\pz\pz &    982.5192 &  0.90 &   F \cr
 17 &  3 &  6 40 47.3 & $-$50 52 51 & 22.95 &  23.20 & 0.25 &  0.06033 &  982.5790 &  0.70 &   F \cr
 18 &  3 &  6 40 44.0 & $-$50 53 13 & 22.74 &  22.97 & 0.23 &  0.05514 &  982.5512 &  0.60 &   O \cr
 19 &  3 &  6 40 42.5 & $-$50 53 15 & 22.63 &  22.90 & 0.27 &  0.06489 &  982.5591 &  0.70 &   O \cr
 20 &  3 &  6 40 33.7 & $-$50 52 11 & 22.84 &  23.12 & 0.28 &  0.06738 &  982.5756 &  0.85 &   F \cr
\end{planotable}
}

\clearpage

%\begin{planotable}
%\tablewidth{35pc}
%\tablecaption{Individual Measurements for the Carina Dwarf Cepheids}
%\end{planotable}
%
\clearpage

\setcounter{table}{3}

\begin{planotable}{cccccc}
\tablewidth{35pc}
\tablecaption{Properties of Other Variable Stars}
\tablehead{
\colhead{SSM} & \colhead{Field} & \colhead{$\langle V\rangle$} & 
 \colhead{$\langle B\rangle$} & \colhead{$\langle B\rangle - \langle V\rangle$}
 & \colhead{$P_{SSM}$} \cr 
 & & & & & (days) }
\startdata
\pz74 & 3 & 20.88 & 21.34 & 0.46 & 0.403 \cr
116   & 1 & 20.91 & 21.49 & 0.58 & 0.680 \cr
176   & 1 & 20.70 & 20.95 & 0.25 & $\ldots$ \cr
177   & 1 & 21.29 & 21.76 & 0.46 & $\ldots$ \cr
\end{planotable}

\clearpage

%\begin{planotable}
%\tablewidth{35pc}
%\tablecaption{Individual Measurements for the Other Carina Variables}
%\end{planotable}
%
%\clearpage

\noindent{\bf Figure 1} -- A color-magnitude diagram of Carina from
the data presented in Paper I.  The solid curved line is an isochrone
from Bertelli et al.  (1994) representing a population with an age of
15 Gyr, a metallicity of [Fe/H] = $-$2.23, a true distance modulus of
20.1 and E$(B-V)$ = 0.025.  The parallelogram encloses the region
where Pop~II dwarf Cepheids (SX~Phe stars) are found in other systems
from the data given in Nemec et al. 1994.

\vskip1em

\noindent{\bf Figure 2} -- 
Light curves of the 20 dwarf Cepheids found in Carina.  For each
star the upper plot shows the $V$-band light curve, and the lower plot 
shows the $B$-band light curve phased with the ephemeris determined from
the $V$-band data.  The period (from Table 2) and pulsation mode of each
individual star is also noted.  `F' represents stars pulsating in the
fundamental mode, `O' represents overtone pulsators, and `?' represents cases
where the mode is uncertain.

\vskip1em

\noindent{\bf Figure 3} -- Charts of the three Carina fields showing
the locations of the newly-discovered dwarf Cepheids (denoted V1-V20)
and the known RR~Lyr stars (SSM~74 and SSM~116) and new RR~Lyr
candidates (RR~176).  The RR~Lyr candidates from Saha et al. (1986)
that we determine are {\it not} variable are denoted by their
identification number from that paper, but with no alphabetic
prefix. SSM~177 is a candidate eclipsing binary as discussed in the
text.  Figures a, b, and c, show fields 1, 2, and 3, respectively.
Note that V14 is the faint star located between the two brighter stars
at its location.  The 1950.0 coordinates of these three fields are
$6^h 39^m 53^s$, $-50^\circ 56' 59''$ for field 1, 
$6^h 40^m 07^s$, $-50^\circ 52' 52''$ for field 2, and
$6^h 40^m 42^s$, $-50^\circ 52' 18''$ for field 3.  
For Field~1, north is up, and east is to the left; for the charts for
Fields 2 and 3, north is to the left, and east is down.  Each field is
3.9 arcmin on a side.  Further details can be found in Paper~I.

\vskip1em

\noindent{\bf Figure 4} --
The color-magnitude diagram of Carina from Figure 1.  The 
square symbols show the locations of the fundamental-mode DCs using the
mean magnitudes and colors from Table 2; the circles denote overtone
pulsators; the triangle is the one star whose mode is uncertain from the
PL relation (Figures 5 and 6).  
The three RR~Lyr stars are shown as bold crosses
using the data from Table 4 and SSM~177, a binary candidate, is shown as
a large open square.  The enclosed region is described in the
text; the horizontal dashed line represents our approximation of the 
faint magnitude limit to which we could identify dwarf Cepheids with 
amplitudes $\gsim$ 0.25 mag.

\vskip1em

\noindent {\bf Figure 5} -- The period-apparent magnitude relation for
the dwarf Cepheids in Carina.  The solid lines correspond to the PLM
relations proposed by NNL for fundamental-mode and overtone dwarf
Cepheids (the lower and upper lines, respectively) using the true
distance modulus (20.09), E$(B-V)$ (0.025) and metallicity for the
younger population ([Fe/H] = $-$2.0) proposed by Smecker-Hane et
al. (1994b).  The dashed lines represent the uncertainty of 0.06 mag
in this distance modulus.  The slopes and relative offsets of the PLM
relations agree well with the Carina observations for these assumed
distance/reddening/metallicity values.

\vskip1em

\noindent {\bf Figure 6} -- The same as Figure 6 except now for the
parameters proposed by Mighell (1997): $(m-M)_0 = 19.89$, E$(B-V) =
0.05$, and [Fe/H] = $-$1.9.  The mean offset of these PL relations from
the data is 0.09 $\pm$ 0.02 mag such that the best-fit distance
modulus is 19.98 $\pm$ 0.02 based on the DCs and the NNL PLM relations
alone.

\vskip1em

\noindent {\bf Figure 7} -- The color-magnitude diagram of Carina from
Figure 1 showing the instability strip boundaries from Breger (1979;
dashed lines) and Mateo (1993; solid lines).

\end{document}